\documentclass[lettersize,journal]{IEEEtran}
\usepackage{amsmath,amsfonts}
\usepackage{algorithmic}
\usepackage{algorithm}
\usepackage{array}
\usepackage[caption=false,font=normalsize,labelfont=sf,textfont=sf]{subfig}
\usepackage{textcomp}
\usepackage{stfloats}
\usepackage{url}
\usepackage{verbatim}
\usepackage{graphicx}
\usepackage{cite}
\usepackage{hyperref}
\usepackage{color}
\begin{document}

\title{Concerns for Self-Localization of Ad-Hoc Arrays Using Time Difference of Arrivals}

\author{Faxian Cao
\thanks{This work was supported by China Scholarship Council}
\thanks{F. Cao is with School of Computer Science, University of Hull, Hull HU6 7RX, U.K. (e-mail: faxian.cao-2022@hull.ac.uk).}}

\markboth{IEEE Transactions on ...}%
{Shell \MakeLowercase{\textit{et al.}}: A Sample Article Using IEEEtran.cls for IEEE Journals}


\maketitle

\begin{abstract}
This document presents  some insights and observations regarding the paper that was published in IEEE Transactions on Signal Processing (TSP), titled "\textit{Self-Localization of Ad-Hoc Arrays Using Time Difference of Arrivals}". 
 In the spirit of constructive feedback, I wish to highlight two key areas of consideration. The first pertains to aspects related to methodology, experimental results, and statements made in the paper. The second part addresses specific equation/typographical errors.
This work aims to initiate a constructive dialogue concerning certain aspects of the paper published in IEEE TSP. Our intention is to provide feedback that contributes to the ongoing improvement of the paper's robustness and clarity. 
\end{abstract}


\section{Introduction}
\IEEEPARstart{T}{his} comment correspondence raises certain concerns regarding the published paper in IEEE Transactions on Signal Processing (TSP), titled "\textit{Self-Localization of Ad-Hoc Arrays Using Time Difference of Arrivals}", authored by \textit{Wang et al.}~\cite{1}. These concerns encompass elements such as a methodology that appears to deviate from the fundamental Herz sampling principle~\cite{2}, some experimental results that seem implausible, and some statements lacking adequate support. Therefore, the primary focus of apprehension for this published paper by \textit{Wang et al.}~\cite{1}
in IEEE TSP revolves around its perceived lack of rigor and clarity.


In the following sections of this comment correspondence, we would like to bring attention to certain aspects of \textit{Wang et al.}'s paper~\cite{1}. Section II discusses the method employed for estimating the time difference of arrival (TDOA) and elaboates why it is deemed confusing. Additionally, there are some concerns regarding experimental results that may appear implausible and statements lacking substantial support.  
Moreover, Section III aims to draw attention to potential equation/typo errors identified in  \textit{Wang et al.}'s paper~\cite{1}.
Finally, the conclusion is drawn in  Section IV.

\section{Concerns for Method, Experimental Results and Statements}

In this section, we would like to highlight the aforementioned concerns observed in 
\textit{Wang et al.}'s paper~\cite{1}.
We begin by addressing concerns related to the method employed for estimating the TDOA.

First, we discuss the Fig. 8 presented in Section VII-D~\cite{1}, \textit{Wang et al.}~\cite{1} simulate the impact of noise on TDOA and its effect on the localization accuracy, it is observed that the \textit{Wang et al.}'s method~\cite{1} performs well when the noise intensity is less than $10^{-6}s$. However, in Section VII-E, when simulating the real environment using a simulated audio signal, to achieve TDOA errors less than $10^{-6}s$, the authors state that they can search the TDOA space with a small step size of $10^{-9}s$ by utilizing $48k$ Hz mics while applying the generalized cross-correlation with phase transform algorithm~\cite{6} for estimating TDOA (refer to the last sentence of the first paragraph on page 12 in \textit{Wang et al.}'s paper~\cite{1}, located to the left of Figure 9 in the same paper). 
However, this approach seems impractical, contradicting the basic physical principles. Specifically, considering a mic sampling rate of $48k$ Hz, the distance between two adjacent samples in received signal is approximately $2.08 \times 10^{-5}s$ since $\frac{1s}{48000} \approx 2.08 \times 10^{-5}s$. Therefore, the minimal searching step is around $2.08 \times 10^{-5}s$, which has a significant difference from what has been stated by the authors~\cite{1}. This discrepancy raises serious concerns. 

Despite this, in Section VII-E~\cite{1}, \textit{Wang et al.}~\cite{1} present TDOA estimation errors, localization errors, and time offset estimation errors in Fig. 9 under varying reverberation times, constituting a realistic simulation. Specifically, in Figure 9(a), the authors illustrate TDOA estimation errors under different reverberation times. It is evident that the median values of TDOA estimation errors are approximately $10^{-7}s$, $6\times 10^{-7}s$, $3\times 10^{-7}s$, $10^{-6}s$ and $9 \times 10^{-5}s$ for reverberation time of $0.1 s$, $0.3s$, $0.5s$, $0.7s$ and $1s$, respectively. This results seem quite inconsistent as the median TDOA estimation errors exhibit minimal change despite an increase in reverberation time, which contradicts. Typically, more reverberation time introduces additional noise to the mic-received signal, leading to increased errors in TDOA measurements. Such a result is available in the literature ~\cite{4} (published in IEEE
TSP) and in line with the general concept. Note that the results produced in ~\cite{4} presents median TDOA estimation errors of approximately $6\times 10^{-7}s$, $7\times 10^{-5}s$, and $8\times 10^{-4}s$ for reverberation times of $0.1s$, $0.3s$, and $0.5s$, respectively. Moreover, the same simulation settings (same mics sampling rate ($48k$ Hz) and
room size (10m × 10m × 3m) ) were observed in both papers. Furthermore, these results from publication \cite{4} are verified and reproducible\footnote{See link: \url{https://github.com/swing-research/xtdoa/tree/master}}. However, these reproducible outcomes in publication \cite{4} significantly differ from the results presented in \textit{Wang et al.}'s paper~\cite{1}.

In summary, \textit{Wang et al.}~\cite{1} propose a novel method that can ignore the basic sampling principle~\cite{2} without providing comprehensive details. Subsequently,  utilzing this method, the TDOA results presented in Fig. 9(a) of \textit{Wang et al.}'s paper~\cite{1} exhibit significant disparities compared to reproducible outcomes in another publication~\cite{4}. 
This raises considerable confusion among readers concerning various experimental results and statements in ~\cite{1}, encompassing Figures 9 and 10 in Sections VII-E and VII-F, the last sentence in Section VII-F, and several statements in the first paragraph of the Conclusion in Section VIII~\cite{1}.

\section{Equation/Typo Errors}
\subsection{Equation errors in Section V-A}
Eqs. (21), (22), (24), (25) and (27)  of Section V-A in \textit{Wang et al.}'s paper~\cite{1} should be
\begin{small}
    \begin{align}
\label{eq1}
    &\frac{d_{1,1}}{c}-\delta_1=0 \Rightarrow \delta_1=\frac{d_{1,1}}{c}=\frac{\|s_1\|}{c}, \\
   \label{eq2}  & \frac{d_{1,j}}{c}-\delta_1+\eta_j=0 \Rightarrow \eta_j=\delta_1- \frac{d_{1,j}}{c}=\frac{\|s_1\|-\|s_j\|}{c}, \\
  \label{eq3}   & \frac{1}{N}\sum_{j=1}^{N}t_{i,j}=\frac{1}{N}\sum_{j=1}^{N}(\frac{d_{i,j}}{c}-\delta_i+\eta_j), \\ 
 \label{eq4}   &  \Rightarrow \delta_i=\frac{1}{N}\sum_{j=1}^{N}(\frac{d_{i,j}}{c}+\eta_j)-\frac{1}{N}\sum_{j=1}^{N}t_{i,j}, \\
  \label{eq5}  &   \frac{1}{N}\sum_{j=1}^{N}(\frac{d_{i,j}-\|s_j\|}{c}-t_{i,j})\leq\delta_i\leq      \frac{1}{N}\sum_{j=1}^{N}(\frac{d_{i,j}+\|s_1\|}{c}-t_{i,j}),
\end{align}
\end{small}
respectively, where $i=2, \cdots, M$ and $j=2, \cdots, N$.





With Eqs. (\ref{eq1}), (\ref{eq3}) and (\ref{eq4}) above, Eq. (28) in \textit{Wang et al.}'s paper~\cite{1} should be
\begin{equation}
\label{eq6}
    \begin{cases}
       0\leq \delta_1\leq \frac{d_{max}}{c} \\
        -\frac{d_{max}}{c}\leq \eta_j\leq \frac{d_{max}}{c} \\
   -\frac{d_{max}}{c}-\frac{1}{N}\sum_{j=1}^{N}t_{i,j}  \leq \delta_i\leq \frac{2d_{max}}{c}-\frac{1}{N}\sum_{j=1}^{N}t_{i,j}
    \end{cases}.
\end{equation}

Upon inspection of Eq. (\ref{eq6}), we can see the statements below Eq. (28) in \textit{Wang et al.}'s paper~\cite{1} are unsubstantiated.

\subsection{Equation errors in Appendix A}
Eq. (A2), (A4), (A5) and (A6) of Appendix A of \textit{Wang et al.}'s paper~\cite{1} in \cite{1} should be
\begin{align}
   & \frac{\partial [U]_{i-1,j-1}}{\partial\eta_k}=(-2(t_{i,j}-t_{1,j})+2(\delta_1-\delta_i))\cdot\uparrow_{j,k},  \\
     &\frac{\partial{[FX-G]}_{i-1,j-1}}{\partial\delta_1}= \sum_{u=1}^{3}(2\eta_{u+1}-2(t_{i,u+1}-t_{1,1}))[X]_{u,j-1}) \nonumber \\
& \qquad  \qquad \qquad \qquad -(2\eta_{j+3}-2(t_{i,j+3}-t_{1,1})), \\
&\frac{\partial{[FX-G]}_{i-1,j-1}}{\partial\delta_k} = \{\sum_{u=1}^{3}2(t_{i,u+1}-t_{i,1}-\eta_{u+1})[X]_{u,j-1} \nonumber \\
       & \qquad \qquad \qquad \qquad +2\eta_{j+3}-(2(t_{i,j+3}-\dot{t}_{i,1}))\} \cdot \uparrow_{i,k},  \\
       & \frac{\partial V_{i-1,j-1}}{\partial\eta_k}=\frac{\partial{(FX-G)}_{i-1,j-1}}{\partial \eta_k} \nonumber \\ 
&=\Bigg\{
     \begin{matrix}
     2(-(t_{i,k}-t_{1,k})+(\delta_1-\delta_i))[X]_{k-1,j-1} & k=2,\cdots,4 \\
     2((t_{i,j+3}-t_{1,j+3})-(\delta_1-\delta_i))\cdot \uparrow_{k,j+3} & k=5,\cdots,N
     \end{matrix},
\end{align}
respectively,
where $i=2, \cdots, M$ and $j=2, \cdots, N$. 





\subsection{Equation errors in Appendix B}
Eq. (B1), (B2), (B4) and (B7) of Appendix B in \textit{Wang et al.}'s paper~\cite{1} should be
\begin{align}
 &   [v_A]_{i-1}=\sum_{u=1}^3 [D_LC]_{i-1,u}^2-2[D_LC]_{i-1,1}s_{x1}- \overline{g}_{i}, \\
&    \frac{\partial [v_A]_{i-1}}{\partial[C]_{k,l}}=2[D_LC]_{i-1,l}[D_L]_{i-1,k}-2\uparrow_{l,1}[D_L]_{i-1,k}s_{x1}, \\
&    [v_B]_{j-1}=\sum_{u=1}^{3}[C^{-1} \Sigma_3 D_R^T]_{u,j-1}^2 +
2[C^{-1} \Sigma_3 D_R^T]_{1,j-1}s_{x1}-\overline{\overline{g}}_{j}, \\
&   \frac{\partial [v_B]_{j-1}}{\partial s_{x1}}=2[C^{-1} \Sigma_3 D_R^T]_{1,j-1},
\end{align}
respectively,
where $i=2, \cdots, M$, $j=2, \cdots, N$, $k=1, 2, 3$ and $l=1, 2, 3$.





\subsection{Other errors}
This subsection shows equation errors at other subsections in \textit{Wang et al.}'s paper~\cite{1}. Eq. (36) in \textit{Wang et al.}'s paper~\cite{1} should be 
\begin{equation}
    \begin{cases}
        \overline{g}_{i,j}=g_{i,j}-g_{1,j}-c^2[\hat{D}]_{i-1,j-1} \\
 \overline{\overline{g}}_{i,j}=g_{i,j}-g_{i,1}-c^2[\hat{D}]_{i-1,j-1} \\
    \end{cases},
\end{equation}
for $i=2,\cdots, M$ and $j=2,\cdots, N$. Besides, in Table II of \textit{Wang et al.}'s paper~\cite{1}, the $v_{T1}$ should be $v_{T2}$ for $10^{-15}$.

\section{Conclusion}
This document provided some serious concerns for \textit{Wang et al.}'s paper~\cite{1}.  Section II first evaluated the method in \textit{Wang et al.}'s paper for searching TDOA violates the basic sampling principle, then compared the unsupported experimental results in \textit{Wang et al.}'s paper with the reproducible results of another publication~\cite{4} in IEEE TSP under the same mics sampling rate and room size, finally, we conclude that some statements from \textit{Wang et al.}'s paper~\cite{1} are confusing since those corresponding statements are based on the unsubstantiated methodology and unsupported experimental results. From Section III, it is obvious that they are many equation/typo errors in publications~\cite{1}. Thus, lacking rigor and clarity is the main problem of this publication~\cite{1}, resulting in the difficulties to evaluate the \textit{real situation} pertaining to the method and experimental results as well as statements in \textit{Wang et al.}'s paper~\cite{1}.


 




\vfill


\begin{thebibliography}{1}
\bibliographystyle{IEEEtran}
\bibitem{1}
L. Wang, T. K. Hon, J. D. Reiss, and A. Cavallaro, "Self-localization of ad-hoc arrays using time difference of arrivals," \textit{IEEE Trans. Signal Process.}, vol. 64, no. 4, pp. 1018-1033, 2015.
\bibitem{2} 
\url{https://en.wikipedia.org/wiki/Sampling_(signal_processing)}
\bibitem{6}
M. S. Brandstein, and H. F. Silverman, "A robust method for speech signal time-delay estimation in reverberant rooms," in {\em Proc. IEEE Int. Conf. Acoust. Speech, Signal Process.}, vol. 1, pp. 375-378, 1997.
\bibitem{4} 
D. E. Badawy, V. Larsson, M. Pollefeys, and I. Dokmanic,  "Localizing unsynchronized sensors with unknown sources," {\em IEEE Trans. Signal Process.}, vol. 71, pp. 641-654, 2023.

\end{thebibliography}
\end{document}